\documentclass[11pt,a4paper]{article}

\usepackage{amsmath}
\usepackage{amsthm,amssymb}

\usepackage{graphicx} 

\newcommand{\beq}{\begin{equation}}
\newcommand{\eeq}{\end{equation}}

\newcommand{\Ai}{\mbox{\rm Ai}}

\numberwithin{equation}{section}

\begin{document}
\title{Pressure exerted by a vesicle on a surface}
\author{A L Owczarek$^1$ and T Prellberg$^2$\\
  \footnotesize
  \begin{minipage}{13cm}
    $^1$ Department of Mathematics and Statistics,\\
    The University of Melbourne, Parkville, Vic 3010, Australia.\\
    \texttt{owczarek@unimelb.edu.au}\\[1ex] 
$^2$ School of Mathematical Sciences\\
Queen Mary University of London\\
Mile End Road, London E1 4NS, UK\\
\texttt{t.prellberg@qmul.ac.uk}
\end{minipage}
}

\maketitle  

\begin{abstract}

Several recent works have considered the pressure exerted on a wall by a model polymer. We extend this
consideration to vesicles attached to a wall, and hence include osmotic pressure. We do this by considering 
a two-dimensional directed model, namely that of area-weighted Dyck paths.

Not surprisingly, the pressure exerted by the vesicle on the wall depends on the osmotic pressure inside,
especially its sign. Here, we discuss the scaling of this pressure in the different regimes, paying particular
attention to the crossover between positive and negative osmotic pressure. In our directed model, there
exists an underlying Airy function scaling form, from which we extract the dependence of the bulk pressure
on small osmotic pressures.

\end{abstract}

\section{Introduction}
A polymer attached to a wall produces a force because of the loss of entropy on the wall. This has been measured experimentally  \cite{bijsterbosch1995a-a,carignano1995a-a,currie2003a-a} and recently described theoretically in two dimensions using lattice walk models \cite{jensen2013a-a,janse2013a-a}. There has also been work concerning the entropic pressure of a polymer in the bulk \cite{gassoumov2013a-a}. Lattice walks and polygons on two dimensional lattices have in the past been utilised to model simple vesicles \cite{leibler1987a-a,brak1990a-a,fisher1991a-a,owczarek1993a-:a,brak1994a-:a}, where there can be an internal pressure. Here we explore the competition between the bulk internal pressure and the point pressure caused by entropy loss when a vesicle is fixed to a wall in two dimensions. Our study involves an exactly solved model of vesicles \cite{owczarek1993a-:a,prellberg1995c-:a}, namely, area-weighted Dyck paths \cite{janse2000a-a,owczarek2010c-:a,owczarek2012b-:a}. 

\begin{figure}[ht]
\begin{center}
\includegraphics[width=0.4\textwidth]{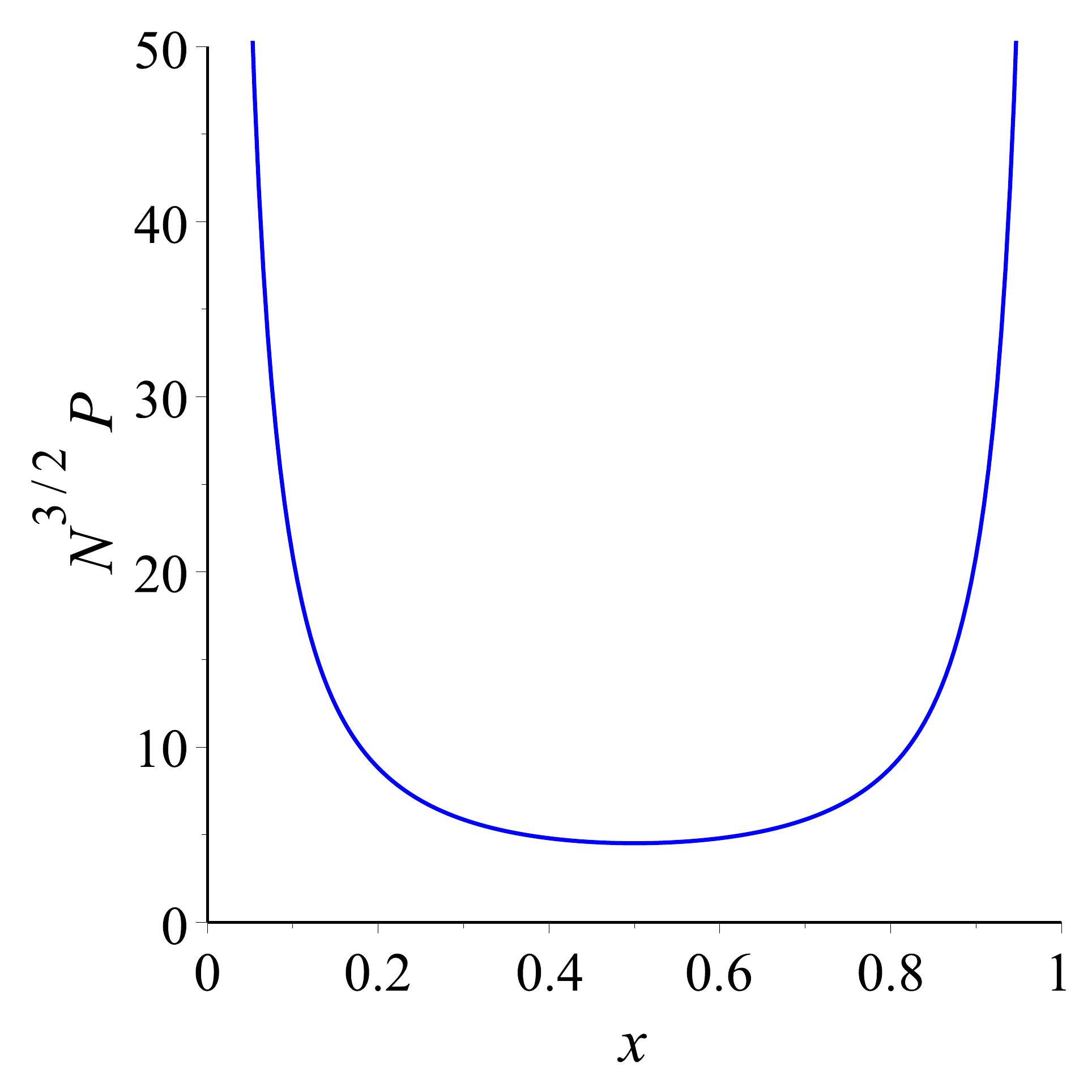}
\end{center}
\caption{Pressure profile for long Dyck paths as a function of the relative distance $x$ from the point of tether.
\label{profile}
}
\end{figure}

Let $Z_N$ be the partition function of some lattice model of rooted configurations of size $N$, for example directed or undirected self-avoiding walks or self-avoiding polygons, and let $Z_N^{(Q)}$ be the partition function conditioned on configurations avoiding a chosen point $Q$ in the lattice. Then the pressure
on the point $Q$ is given by the difference of the finite-size free energies $-\log Z_N$ and $-\log Z_N^{(Q)}$, that is
\beq
\label{generalpressure}
P_N^{(Q)}=-\log Z_N^{(Q)}+\log Z_N\;.
\eeq
Here we take $k_BT=1$ for convenience.
When the configurations are Dyck paths, which are directed paths above the diagonal of a square lattice starting at the origin and ending on the diagonal, this model was analysed in \cite{janse2013a-a}. The pressure at the point $Q=(m,m)$ for walks of length $2N$ is given exactly as
\beq
\label{catalanpressure}
P_N^{(m)}=-\log\left(1-\frac{C_mC_{N-m}}{C_N}\right)\;,
\eeq
where $C_k=\frac1{k+1}\binom{2k}k$ is the $k$-th Catalan number counting $2k$-step Dyck paths. For $N$ and $m$ large, this leads to
\beq
\label{Dyckpathprofile}
P_N^{(m)}=\frac1{\sqrt{\pi x^3(1-x)^3}}\cdot\frac1{N^{3/2}}+O(N^{-5/2})\;,
\eeq
where $x=m/N$ measures the relative distance of the point $Q$ from the origin with respect to the length of the walk. That is, the pressure of the Dyck path
decays to zero as $N^{-3/2}$ in the centre of the Dyck path, with an $x$-dependent profile, as shown in Figure \ref{profile}. 
In contrast, near the boundary the pressure tends to an $N$-independent limiting
value
\beq
P_N^{(m)}\rightarrow-\log\left(1-\frac{C_m}{4^m}\right)\;.
\eeq

\section{The model}

\begin{figure}[ht]
\begin{center}
\includegraphics[width=0.9\textwidth]{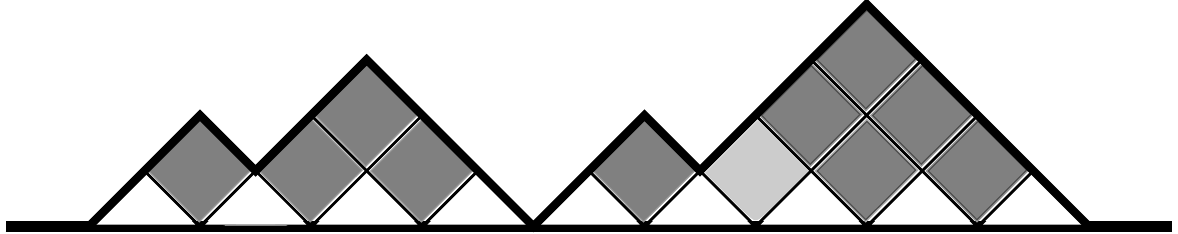}
\end{center}
\caption{A vesicle of length 9 given by a Dyck path with 18 steps, enclosing the 11 shaded plaquettes and having weight $q^{11}$. 
When considering the modified configurations that avoid the lighter coloured
plaquette, which is at distance $m=6$, the vesicle is considered to enclose 10 plaquettes and have weight $q^{10}$.
\label{vesicle}
}
\end{figure}

In an extension of the work described in the introduction, the configurations of the model studied here are area-weighted Dyck paths, where we have
rotated the lattice by $45^{\circ}$ for convenience as shown in Figure \ref{vesicle}. To be more precise, we weight each full square plaquette between the Dyck path and the surface 
with a weight $q=\exp(\Pi)$, where $\Pi$ is the osmotic pressure. 

We use these paths to model vesicles adsorbed at the surface, that is to say that we consider them as vesicles with the bottom part of the membrane 
firmly attached to the surface. As described above, to calculate the pressure we need to consider a slightly modified set of configurations that avoid some point.
For our vesicle model this means that the bottom of the vesicle does not include a particular plaquette, see Figure \ref{vesicle}.

Denoting the set of all $2N$-step Dyck paths by ${\cal D}_N$, the partition function for unrestricted vesicles of length $N$ is given by
\beq
Z_N(q)=\sum_{\varphi\in{\cal D}_N}q^{A(\varphi)}\;,
\eeq
where $A(\varphi)$ is the number of plaquettes enclosed by the configuration $\varphi$.

Similarly, denoting the set of all $2N$-step Dyck paths enclosing the surface plaquette at distance $1\leq m\leq N-1$ by ${\cal D}_N^{(m)}\subset{\cal D}_N$, 
the partition function for the restricted vesicles is given by
\beq
Z_N^{(m)}(q)=\sum_{\varphi\in{\cal D}_N^{(m)}}q^{A(\varphi)-1}\;.
\eeq
Note that the distance $m$ is measured as the number of half-plaquettes along the surface to the point of interest.

The configurations in ${\cal D}_N\setminus{\cal D}_N^{(m)}$ are precisely the ones that touch the surface at distance $m$,
whence
\beq
Z_N(q)-qZ_N^{(m)}(q)=\sum_{\varphi\in{\cal D}_N\setminus{\cal D}_N^{(m)}}q^{A(\varphi)}=Z_m(q)Z_{N-m}(q)\;.
\eeq
For $q=1$ this reduces to Equation (\ref{catalanpressure}).
Computing the pressure using Eqn.~(\ref{generalpressure}), we find that the pressure $P_N^{(m)}(q)$ of a Dyck vesicle of length $N$ on the surface at 
distance $m$ is given by
\beq
\label{pressureformula}
P_N^{(m)}(q)=-\log\left(1-\frac{Z_m(q)Z_{N-m}(q)}{Z_N(q)}\right)+\log q\;.
\eeq 

\section{Results}

\subsection{Exact results}

One can recursively calculate $Z_N(q)$ using
\beq
\label{recurrence}
Z_0(q)=1\;,\quad
Z_{N+1}(q)=\sum\limits_{k=0}^Nq^kZ_k(q)Z_{N-k}(q)\;,\quad N\geq 0\;,
\eeq
which for $q=1$ reduces to a well-known recursion for Catalan numbers. One sees that $Z_N(q)$, and therefore also $P_N^{(m)}(q)$, 
is computable in polynomial time. There are no closed-form expressions known. 

However, there are well-known closed-form expressions \cite[Example V.9]{flajolet2009a-a} for the generating function 
\beq
G(z,q)=\sum_{N=0}^\infty Z_N(q)z^N\;,
\eeq
namely,
\beq
G(z,q)=\frac{A_q(z)}{A_q(z/q)},
\eeq
where
\beq
A_q(z)=\sum_{n=0}^\infty\frac{q^{n^2}(-z)^n}{(q;q)_n}
\eeq
is Ramanujan's Airy function. 
Here, we use the $q$-product notation $(t;q)_n=\prod_{k=0}^{n-1}(1-tq^k)$.  

Note that the radius of convergence $z_c(q)$ of $G(z,q)$ is simply related to the thermodynamic limit of the partition function via
\beq
\log z_c(q)=-\lim_{N\to\infty}\frac1N\log Z_N(q)\;.
\eeq
For $q=1$, $G(z,q)$ is simply the Catalan generating function and hence $z_c(1)=1/4$.

At the level of the generating function, the recurrence (\ref{recurrence}) is equivalent to the
functional equation
\beq
\label{functeqn}
G(z,q)=1+zG(z,q)G(qz,q)\;,
\eeq
which in turn leads to a nice continued fraction representation for $G(z,q)$:
\beq
\label{confrac}
G(z,q)=\cfrac1{1-\cfrac z{1-\cfrac{qz}{1-\cfrac{q^2z}{1-\cfrac{q^3z}{1-\ldots}}}}}\;.
\eeq
We can calculate the surface pressure in the thermodynamic limit from the generating function by relating the pressure to 
the density of contacts with the surface. In order to do so, we need to introduce a surface weight $\kappa$ for contacts with the
surface, which we will set to one after the calculation. Let $G(z,q;\kappa)$ be the generating function for area-weighted Dyck paths
where each contact with the surface (except for the origin) is associated with a weight $\kappa$. A simple necklace argument gives
\beq
\label{kappaeqn}
G(z,q;\kappa)=
\frac1{1-z\kappa G(qz,q)}\;.
\eeq
The density of contacts with the surface well away from the ends of the walk is
\beq
\label{rhoeqn}
\rho(q)=-\left.\frac{\partial\log z_c(q;\kappa)}{\partial\log\kappa}\right|_{\kappa=1}\;,
\eeq
where $z_c(q;\kappa)$ is the location of the closest singularity of $G(z,q;\kappa)$ in $z$ to the origin. The surface pressure 
in the thermodynamic limit, for any fixed value of $0<x=m/N<1$, is then given as the constant
\beq
P(q)=-\log(1-\rho(q))+\log q\;.
\eeq
We shall refer to this as the bulk pressure.

\begin{figure}[ht]
\begin{center}
\includegraphics[width=0.5\textwidth]{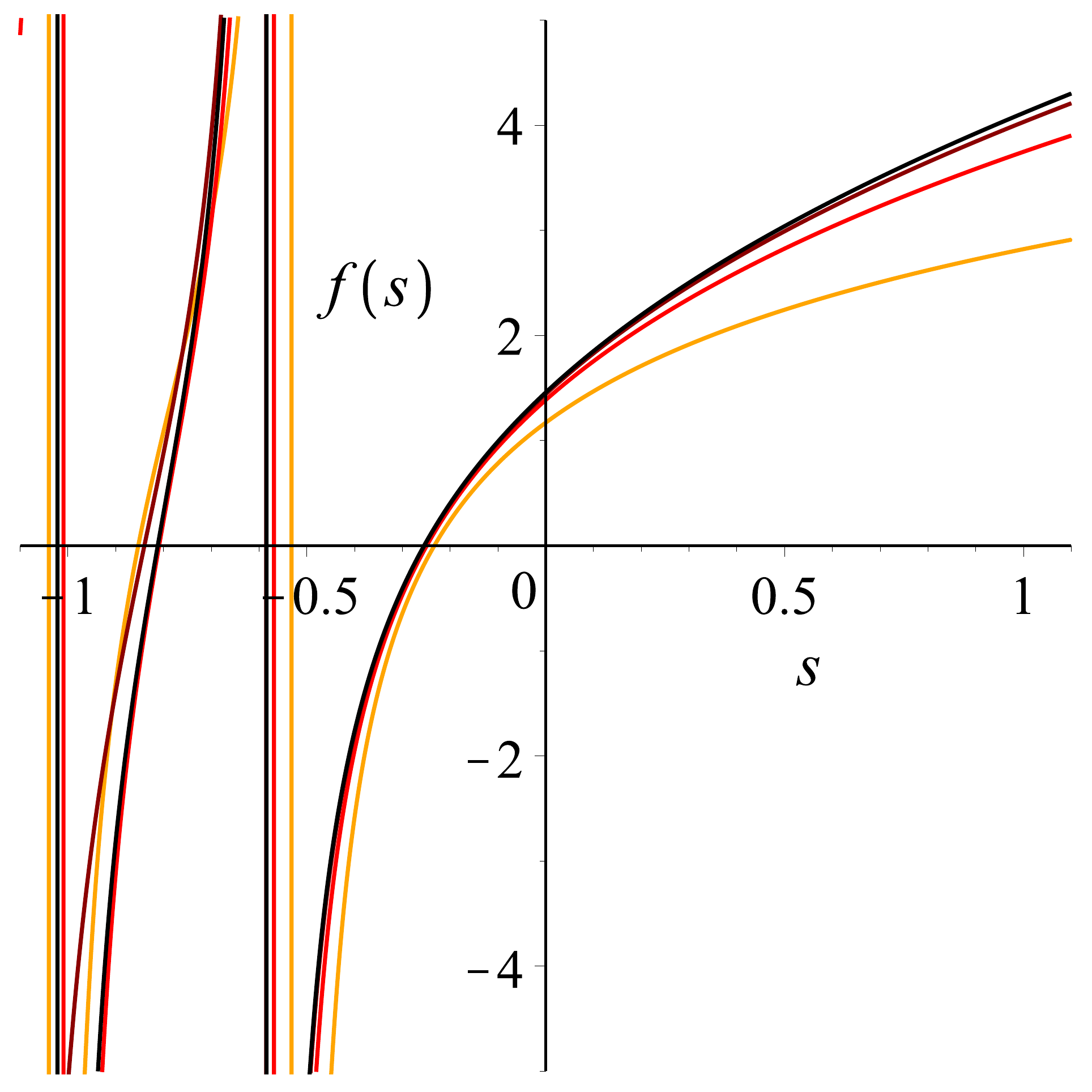}
\end{center}
\caption{Convergence of the scaled generating function $G(z,q)$ to the scaling function $f(s)$ for $q=0.99$, $0.9999$, and $0.999999$. Numerical evaluation
of $G(z,q)$ has been done using the continued fraction representation (\ref{confrac}).
\label{Generatingfunctionscaling}
}
\end{figure}

Much work has been done on the computation of the asymptotics for $q$-series such as those involved here \cite{prellberg1994a-a}.
In the vicinity of $q=1$ and $z=z_c(1)=1/4$, one can show convergence of suitably scaled generating function. More precisely, one finds that
the limit
\beq
\label{scaling}
f(s)=\lim_{q\to1^-}-\left((1-q)^{-1/3}G(1/4-s(1-q)^{2/3},q)-2\right)
\eeq
exists and is equal to the scaling function
\beq
f(s)=-2\frac{\Ai'(4s)}{\Ai(4s)}
\eeq
The convergence to the scaling function is shown in Figure \ref{Generatingfunctionscaling}.

\subsection{Pressure profiles}

\begin{figure}[ht]
\begin{center}
\includegraphics[width=0.3\textwidth]{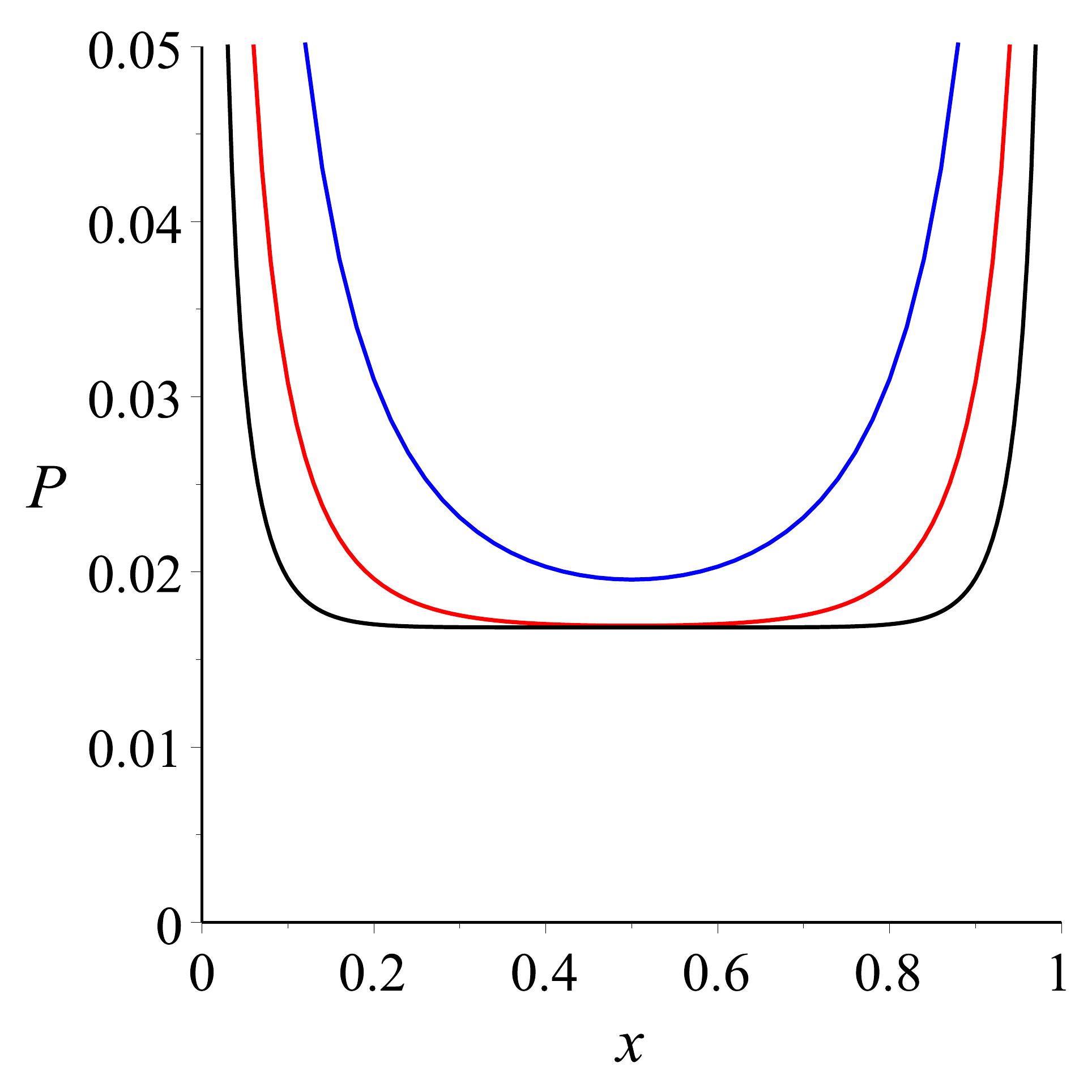}
\includegraphics[width=0.3\textwidth]{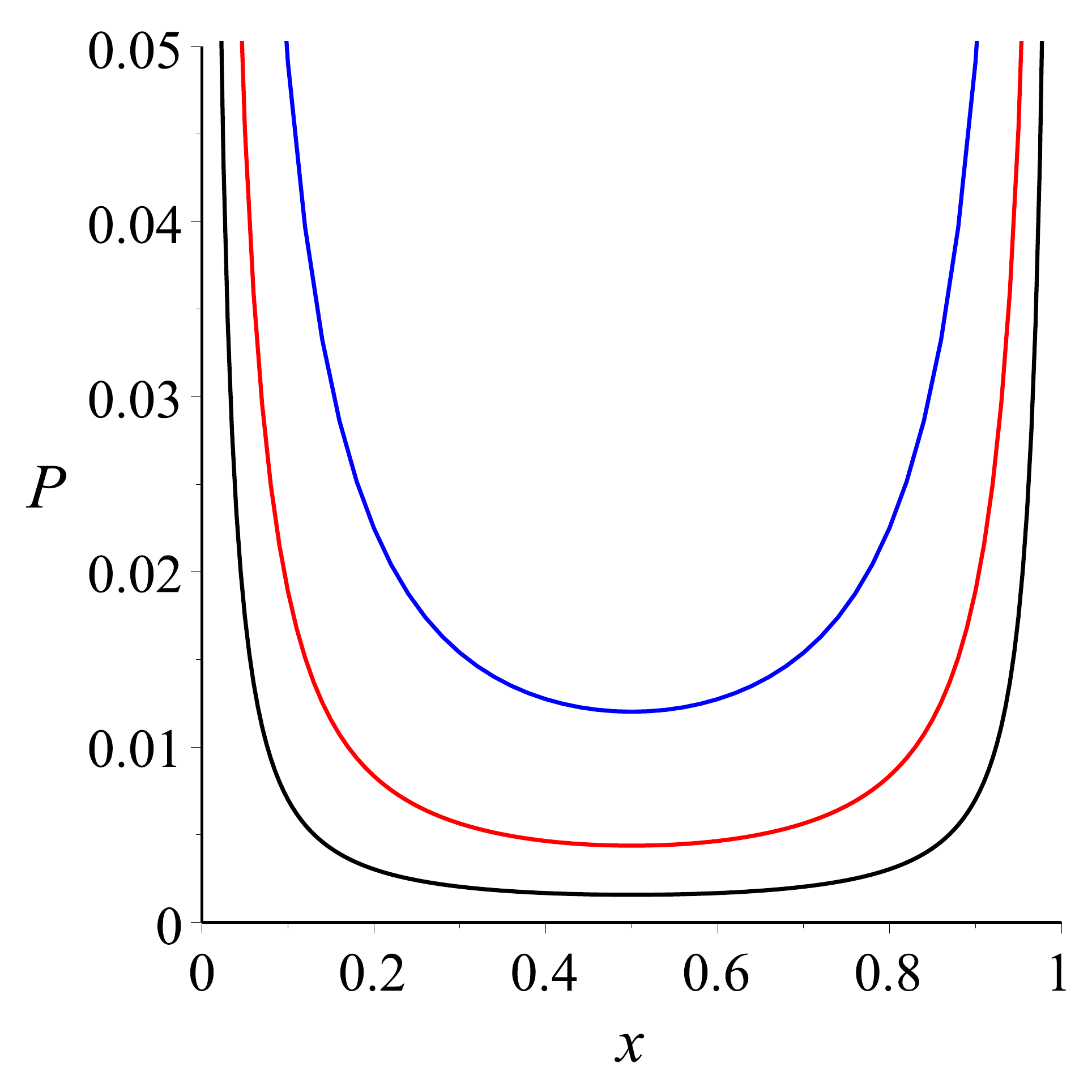}
\includegraphics[width=0.3\textwidth]{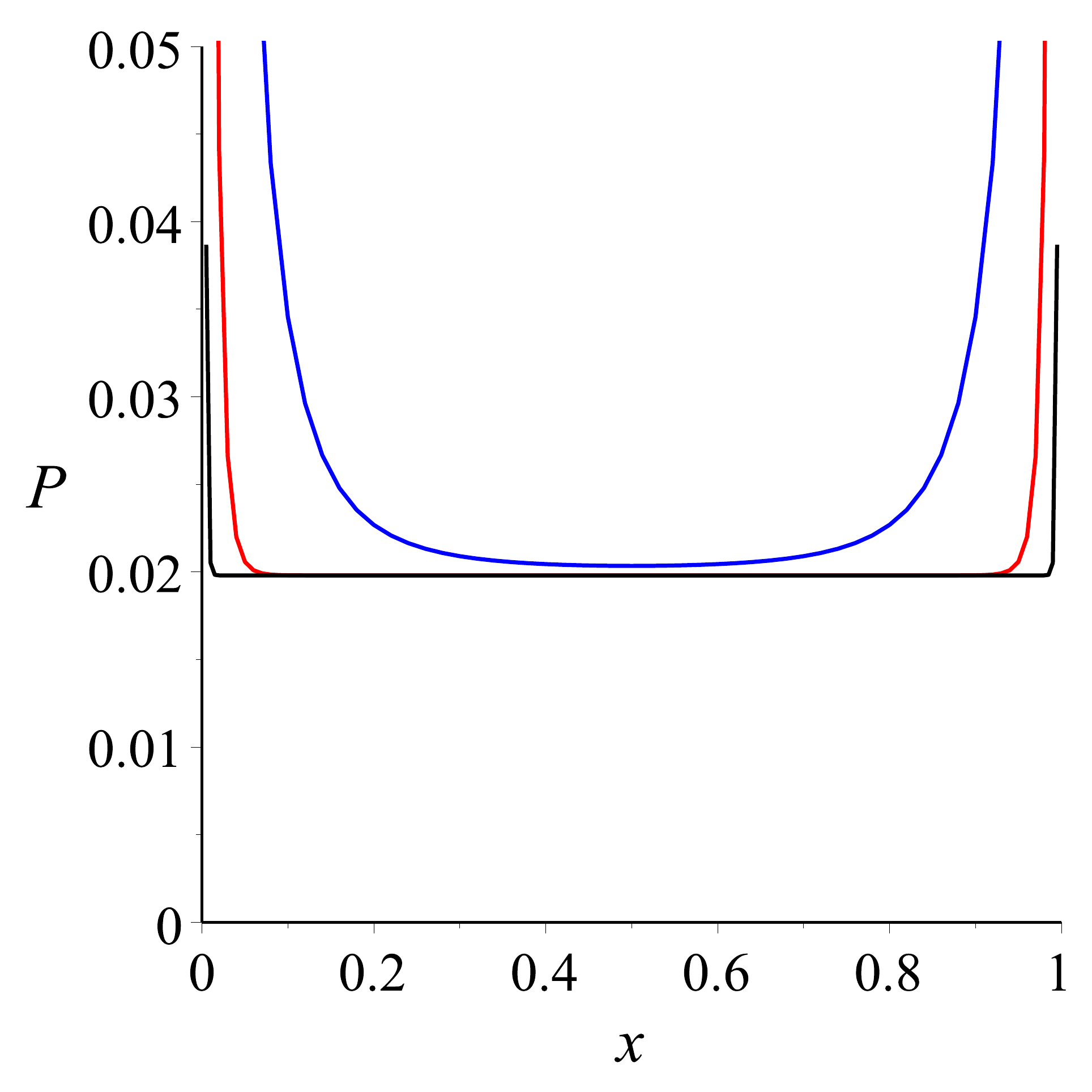}
\end{center}
\caption{Pressure profiles for $q=0.98$ (left), $1.00$ (center), and $1.02$ (right) and lengths $N=50$, $100$, and $200$, from top to bottom.
\label{pressureprofiles}
}
\end{figure}

We now consider the profile of the pressure for finite $N$. In Figure \ref{pressureprofiles} we show 
pressure profiles as a function of $x=m/N$ for three different values of $q$ with positive, zero, and negative
osmotic pressures, and for lengths $N=50$, $100$, and $200$.

\begin{figure}[ht]
\begin{center}
\includegraphics[width=0.3\textwidth]{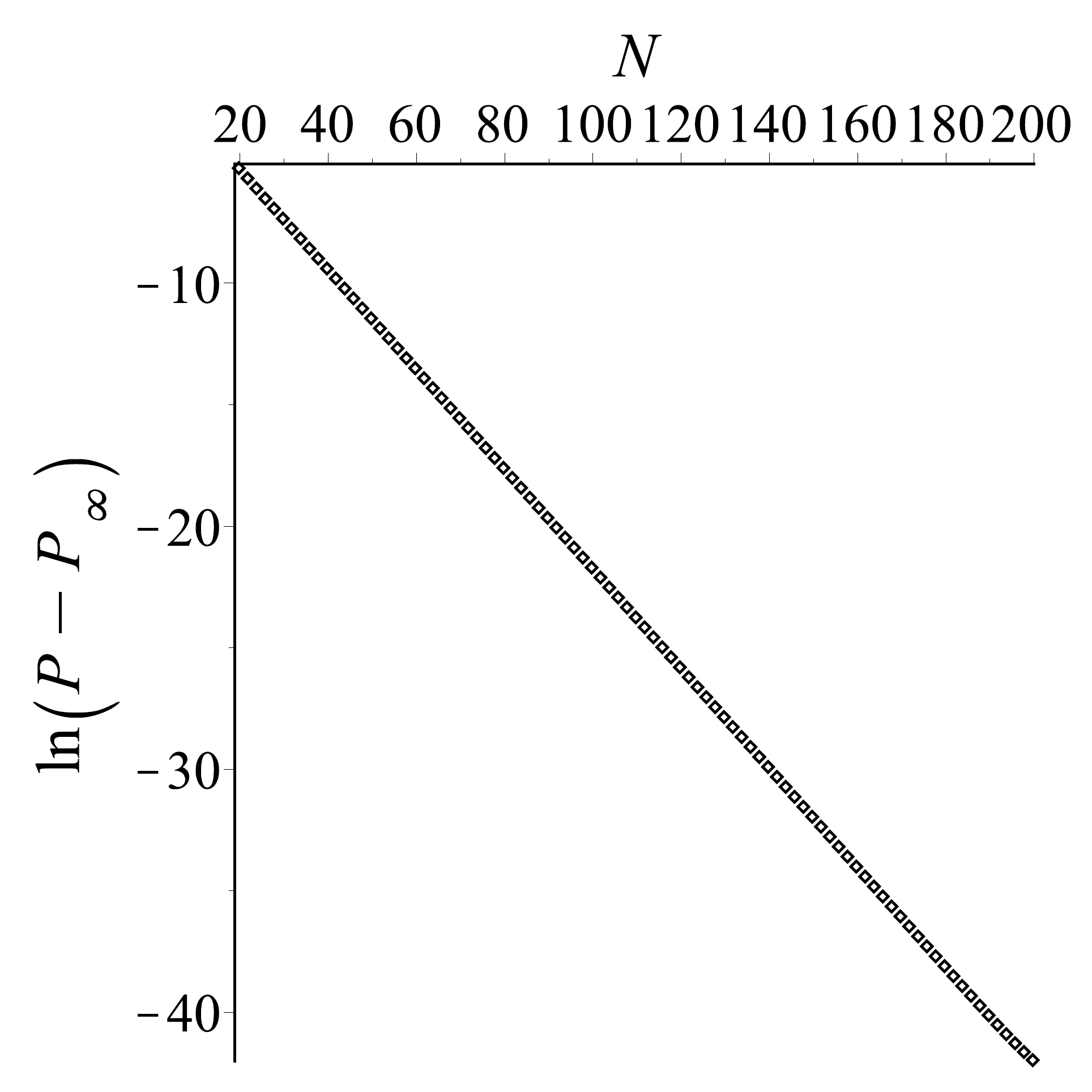}
\includegraphics[width=0.3\textwidth]{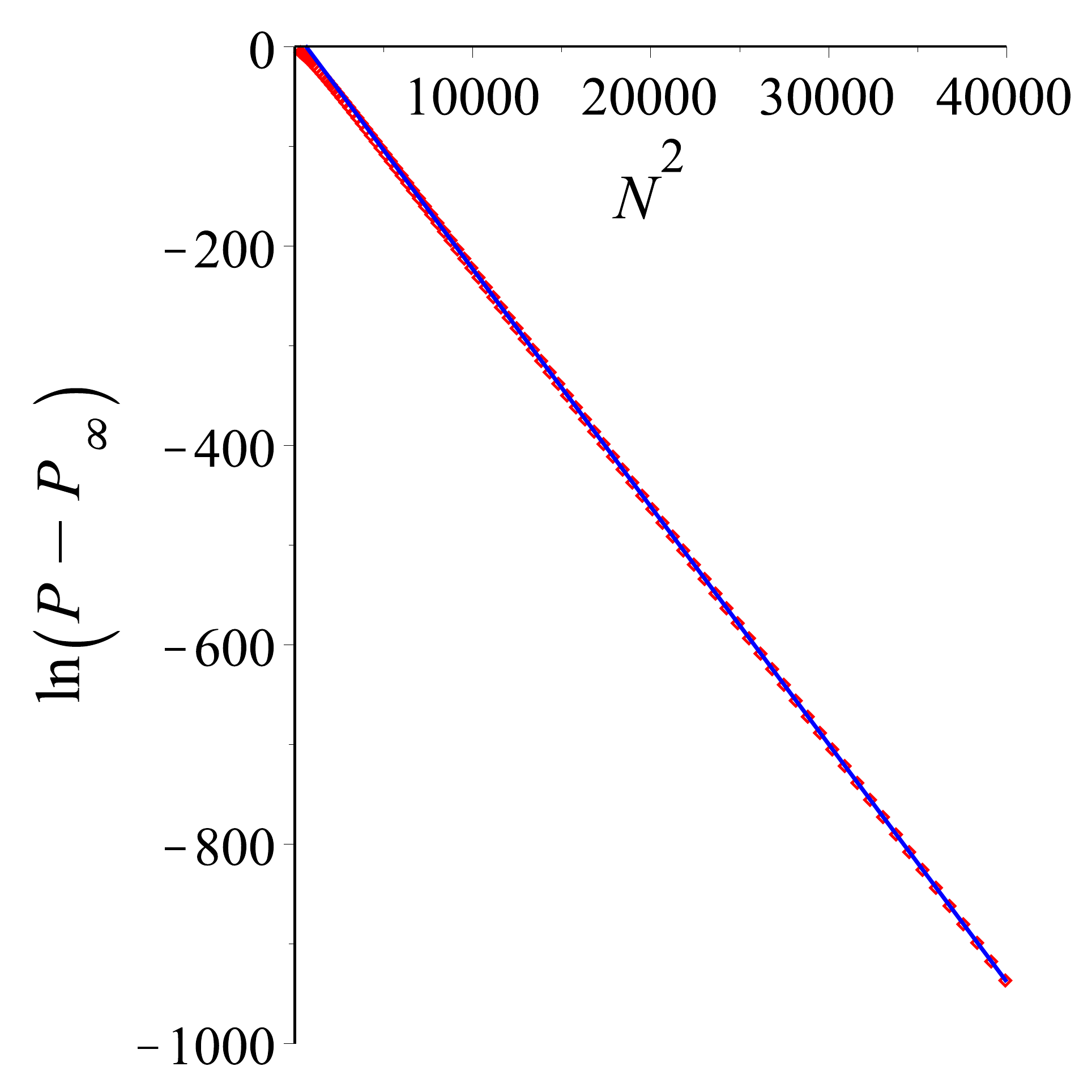}
\end{center}
\caption{The rates of approach to the bulk pressure, calculated via the pressure at the centre of the vesicle for $q=0.9$ (left) and $q=1.1$ (right). Clearly
the convergence is exponential in $N$ for $q<1$ and exponential in $N^2$ for $q>1$. The straight line (right) is our theoretical prediction.
\label{pressuredecay}
}
\end{figure}

We see that the pressure converges to a non-zero bulk pressure when the osmotic pressure is non-zero. For $q=1$, one can compare the finite-size data
shown in this figure to the exactly known scaled profile shown in Figure \ref{profile}.  Moreover, closer inspection shows that the rate of 
convergence is significantly different for positive and negative osmotic pressure. As we know, for $q=1$ the bulk pressure is zero, and convergence obeys
the power law $N^{-3/2}$. For $q\neq1$, the rate of convergence to the bulk pressure is exponential in $N$ and $N^2$ for $q<1$ and $q>1$, respectively,
as can be seen in Figure \ref{pressuredecay}.

One can derive these results in the following way. Convergence to the thermodynamic limit is encoded in the singularity structure of the generating function.
For $q<1$, the leading singularity of the generating function is an isolated simple pole, and the finite-size corrections to scaling are therefore exponential,
with the rate of convergence given by the ratio between the magnitudes of the leading singularity and the sub-leading one. A closer analysis reveals that
the rate depends on the value $x=m/N$ as 
\beq
P_N^{(m)}(q)\sim-\log(1-\rho(q))+\log q+O(e^{-D\min(x,1-x)N})\;,
\eeq
where $D$ does not depend on the value of $x$. The singularity structure can also be seen in Figure \ref{Generatingfunctionscaling} for negative 
values of the scaling parameter $s$.

For $q>1$, it is easier to argue directly via the partition functions. For positive osmotic pressure, configurations with large area dominate, and using arguments
in \cite{prellberg1999a-:a} one can deduce that 
\beq
Z_N(q)\sim\frac1{(q^{-1};q^{-1})_\infty}q^{\binom N2}
\eeq
where $(t;q)_\infty=\prod_{k=0}^\infty(1-tq^k)$ is again a $q$-product. The appearance of the factor $1/(q^{-1};q^{-1})_\infty$ is due to fluctuations around configurations
of maximal area.

Substituting this into Eqn.~(\ref{pressureformula}) shows that
\beq
P_N^{(m)}(q)\sim\log q+\frac1{(q^{-1};q^{-1})_\infty}q^{-m(N-m)}=\log q+\frac1{(q^{-1};q^{-1})_\infty}q^{-x(1-x)N^2}\;.
\eeq
Note that the decay rate depends on the value of $x=m/N$.

\subsection{Bulk pressure}

\begin{figure}[ht]
\begin{center}
\includegraphics[width=0.4\textwidth]{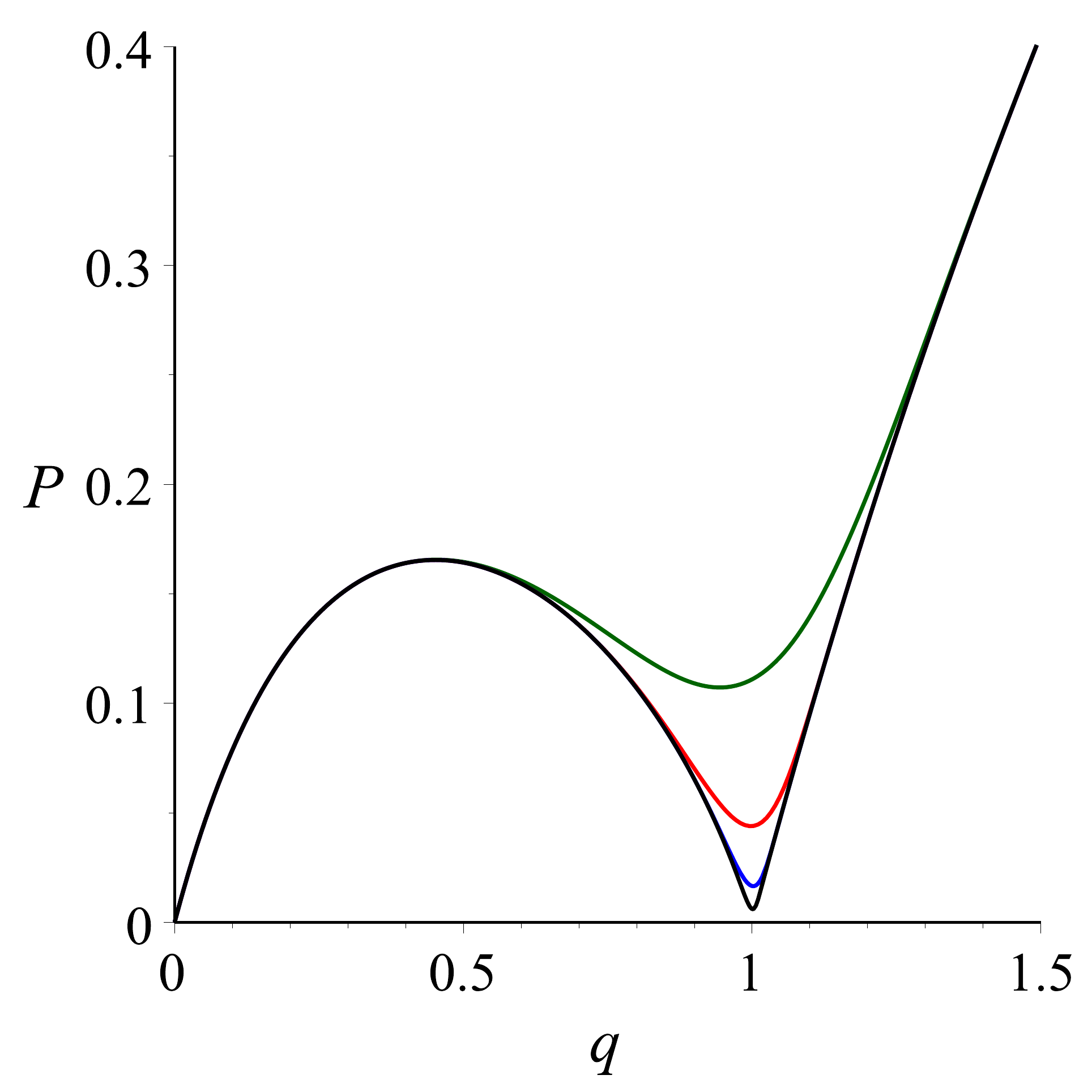}
\includegraphics[width=0.4\textwidth]{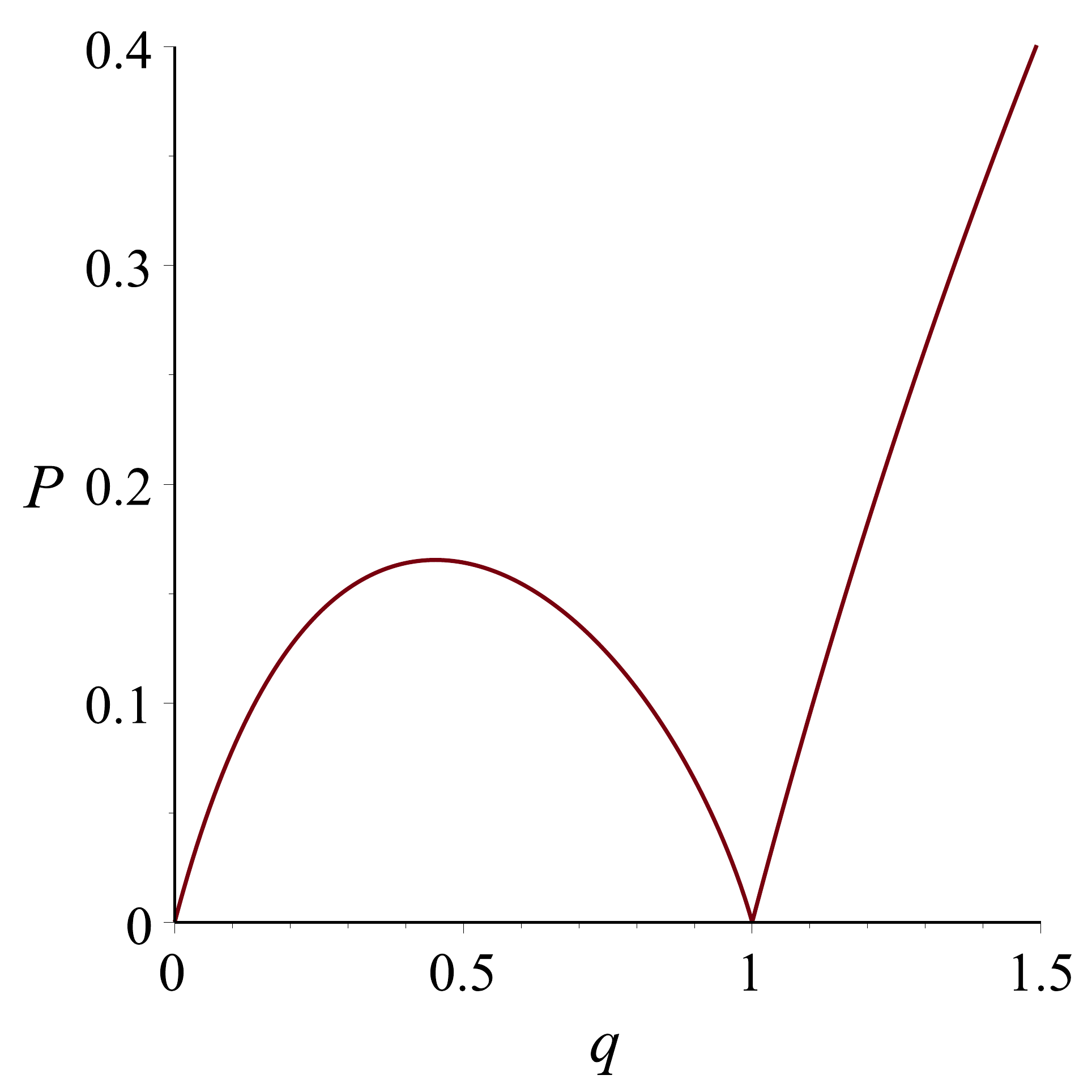}
\end{center}
\caption{A plot of the pressure at the centre of the vesicle against $q$ for vesicle lengths $10$, $20$, $40$ and $80$, from top to bottom (left) and of the thermodynamic limit bulk pressure of the vesicle against $q$ (right).
\label{Pvsq}
}
\end{figure}

We now turn to the consideration of the pressure in the centre of the vesicle, so as to consider the bulk pressure. Using our results above, we already know
that for $q>1$ the bulk pressure is given by $\log q$. In Figure \ref{Pvsq} on the left we have used the recursion (\ref{recurrence}) to calculate the pressure at
the centre of the vesicle as a function of $q$ for $N=10$, $20$, $40$, and $80$. One can see that the convergence to a limit is slowest around $q=1$, which of
course aligns with our predictions above for the rates of convergence in the different regimes.

We have now used the continued fraction expansion (\ref{confrac}) to numerically estimate the thermodynamic limit bulk pressure as a function of $q$, shown on the right in Figure \ref{Pvsq}.
It should be clear that the finite-size curves approach the curve shown here in the limit of large $N$.  It is interesting to see the competition of the osmotic pressure and the entropic pressure for $q<1$. 

Clearly the limiting behaviour of the pressure for $q>1$ is $P(q)\sim q-1$ as $q\to1^+$. Extracting the limiting behaviour of the pressure for $q<1$ is considerably more difficult. 
A calculation starts by considering that from Equations (\ref{functeqn}) and (\ref{kappaeqn}) we find that the critical fugacity $z_c=z_c(q;\kappa)$ satisfies
\beq
G(z_c,q)=\frac{\kappa}{\kappa-1}\;.
\eeq
Differentiating this expression with respect to $\kappa$ allows us to write the density $\rho(q)$ in terms of the generating function as
\beq
\rho(q)=\frac{G(z_c,q)(G(z_c,q)-1)}{z\frac{\partial}{\partial z_c}G(z_c,q)}\;.
\eeq
Utilising the scaling form (\ref{scaling}) now shows that $\rho(q)\sim2(1-q)$ as $q\to1^-$ and hence $P(q)\sim 1-q$. Put together, this implies that
\beq
\label{Pasy}
P(q)\sim|1-q|\quad\text{as $q\to 1$.}
\eeq

\begin{figure}[ht]
\begin{center}
\includegraphics[width=0.4\textwidth]{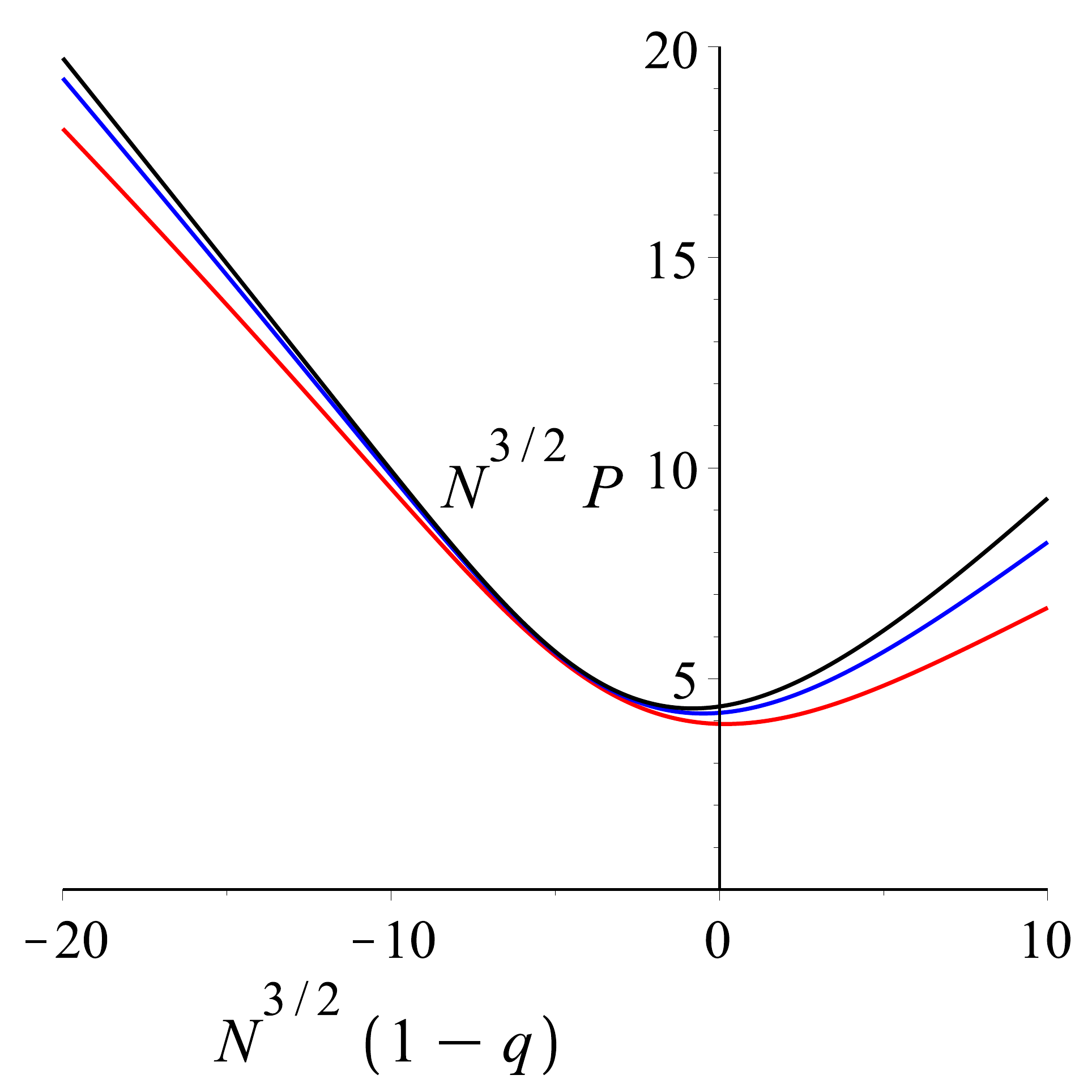}
\end{center}
\caption{An estimate of the finite-size scaling function for the pressure, using $N=40$, $80$, and $160$, from bottom to top.
\label{pressurescaling}
}
\end{figure}

The existence of the scaling function $f(s)$ in the variable $s=(1/4-z)/(1-q)^{2/3}$ implies by standard Laplace transform that there should be a finite-size scaling form for the
pressure in the variable $t=N^{3/2}(1-q)$. Hence we define the scaling function
\beq
g(t)=\lim\limits_{N\to\infty}N^{3/2}P_N^{(N/2)}(1-tN^{-3/2})\;.
\eeq
In Figure \ref{pressurescaling}, we numerically estimate the scaling function $g(t)$ for a range of $t$. Asymptotic matching with (\ref{Pasy}) requires that $g(t)\to|t|$ for $t\to\pm\infty$.
We note the convergence to the scaling function is poor for $q>1$, which arises
because of the unusually differing rates of convergence to the thermodynamic limit in the two regimes.
 
\section*{Acknowledgements}

Financial support from the Australian Research Council via its support
for the Centre of Excellence for Mathematics and Statistics of Complex
Systems is gratefully acknowledged by the authors. A L Owczarek thanks the
School of Mathematical Sciences, Queen Mary, University of London for
hospitality.

\providecommand{\newblock}{}

\end{document}